# Development and validation of a targeted LC-MS/MS quantitation method to monitor cell culture expression of tetanus neurotoxin during vaccine production

Antoine Francotte [a,b], Raphael Esson [c], Eric Abachin [c], Melissa Vanhamme [b], Alexandre Dobly [a], Bruce Carpick [d], Sylvie Uhlrich [c], Jean-François Dierick [e], Celine Vanhee [b*]

[a] Department of Expertise and service provision, quality of vaccines and blood products, Sciensano, 14 rue Juliette Wytsman, 1050 Brussels, Belgium,

[b] Department of Chemical and physical health risks, Medicines and health care products, Sciensano, 14 rue Juliette Wytsman, 1050 Brussels, Belgium,

[c] Sanofi Pasteur, 1541 Avenue Marcel Mérieux, 69280, Marcy l'Etoile, France.

[d] Sanofi Pasteur, 1755 Steeles Ave West, Toronto, Ontario, Canada.

[e] GSK Vaccines, Parc de la Noire Epine, Avenue Flemming 20, B-1300 Wavre, Belgium

*corresponding author:

E-mail address: celine.vanhee@sciensano.be

Postal address: rue Juliette Wytsman, 14 - 1050 Brussels

Telephone: +32 2 642 5142 | Fax: +32 2 642 53 27

## ABSTRACT

The tetanus neurotoxin (TeNT) is one of the most toxic proteins known to man, which prior to the use of the vaccine against the TeNT producing bacteria *Clostridium tetani*, resulted in a 20 % mortality rate upon infection. The clinical detrimental effects of tetanus have decreased immensely since the introduction of global vaccination programs, which depend on sustainable vaccine production. One of the major critical points in the manufacturing of these vaccines is the stable and reproducible production of high levels of toxin by the bacterial seed strains. In order to minimize time loss, the amount of TeNT is often monitored during and at the end of the bacterial culturing. The different methods that are currently available to assess the amount of TeNT in the bacterial medium suffer from variability, lack of sensitivity, and/or require specific antibodies. In accordance with the consistency approach and the three Rs (3Rs), both aiming to reduce the use of animals for testing, in-process monitoring of TeNT production could benefit from animal and antibody-free analytical tools.

In this paper, we describe the development and validation of a new and reliable antibody free targeted LC-MS/MS method that is able to identify and quantify the amount of TeNT present in the bacterial medium during the different production time points up to the harvesting of the TeNT just prior to further upstream purification and detoxification. The quantitation method, validated according to ICH guidelines and by the application of the total error approach, was utilized to assess the amount of TeNT present in the cell culture medium of two TeNT production batches during different steps in the vaccine production process prior to the generation of the toxoid. The amount of TeNT generated under different physical stress conditions applied during bacterial culture was also monitored.

## INTRODUCTION

The tetanus neurotoxin (TeNT) is one of the most toxic proteins known to man, second only to the group of botulinum neurotoxins [1]. TeNT is produced by the widespread Gram-positive soil bacteria *Clostridium tetani* as a single polypeptide (approximately 150 kDa) which is subsequently cleaved into a two-chain active holotoxin, consisting of an N-terminal light chain of 50 kDa and a C-terminal Heavy chain of approximately 100 kDa, linked by a single disulphide bond [2]. The heavy chain is involved in target recognition, binding, and subsequent cell entry while the light chain includes a zinc-dependent endopeptidase that is responsible for the inhibition of neurotransmission [3,4]. This inhibition is responsible for the paralytic clinical effect of tetanus that still causes morbidity and deaths worldwide, particularly in infants and non-immunized adults, despite global vaccination programs [5, 6].

Vaccination against tetanus was introduced in Europe and the United States in the 1920s-1930s and in many countries general vaccination has been performed since the 1940s [7]. The vaccine antigen consists of the chemically detoxified TeNT, termed tetanus toxoid. The vaccine manufacturing process starts with the generation of stable and GMP monitored *C. tetani* seed lots that produce TeNT. Next, the toxin is purified and inactivated by treatment with formaldehyde, generating the toxoid. The toxoid is then mixed with adjuvants, generally aluminum salts, prior to filling and packaging of the vaccine [8]. This final product is then subjected to batch release, an extensive quality control testing performed by both the producer and an official control authority. The latter quality control assays are mainly based on monographs from the different pharmacopeias, including the European Pharmacopoeia (Ph. Eur), the United States Pharmacopoeia (USP), or alternatively World Health Organization (WHO) guidelines [9]. Similarly at the producers end, there is a global interest to apply the consistency approach which relies on the strict application of a quality system and of a consistent production of batches and should result in the reduction in the number of animals used in testing of the biological products [10]. Consequently, this also incorporates a thorough characterization of the product during development and during the production process or, alternatively, in case of changes into the manufacturing process.

One of the major critical points in the production of clostridial vaccines is the stable and reproducible production of high levels of toxin as the industrial TeNT production is sometimes hampered by low titers and occasional batch failures [9,11]. Quantitation of TeNT is traditionally carried out by an *in vitro* flocculation method which expresses the amount of toxin in terms of Limit of flocculation (Lf) units [12-15]. The Lf method is based on binding of the tetanus antigen and a specific antiserum, which produces a flocculent precipitate. However, the Lf test is known to suffer from considerable variation in output and is, per design, semi-quantitative [16, 17]. As an alternative *in vitro* approach to TeNT quantitation, various enzyme-linked immunosorbent assays (ELISA) have been developed [18-22]. Although these ELISAs have proven to be highly specific and sensitive, they rely on the generation and management of either polyclonal antibodies derived directly from the serum of immunized animals, or monoclonal antibodies derived from hybridomas which also require animal immunization [20-25]. Consequently, in accordance with the consistency approach and 3Rs principles, TeNT quantification for in-process product monitoring would benefit from animal- and antibody-free analytical tools. Targeted quantitative LC-MS/MS represents an alternative, animal friendly approach. In recent years, LC-MS methodologies have been successfully applied in the field of vaccine development [26, 27].

In this paper, we describe the development and validation of a new and robust antibody free targeted LC-MS/MS method that is able to identify and quantify the amount of TeNT present in the bacterial medium at different production time points up to the harvesting of the TeNT just prior to further upstream purification and detoxification. Our data also demonstrate that the method is suitable for real-life in-process monitoring of TeNT production and could replace the currently utilized semi-quantitative flocculation assay, known to suffer from lack of sensitivity and robustness. Additionally, our data also demonstrate that this newly developed method can also be utilized to assess the impact of changes in culturing conditions, known to affect the classical flocculation test, or to support the optimization of culture process. This work was carried out as part of the Innovative Medicines Initiative 2 (IMI2) project VAC2VAC (Vaccine batch to vaccine batch comparison by consistency testing).

## MATERIALS AND METHODS

### 2.1. Culture supernatant production

2.1.1 Bacterial strains and standard culturing conditions

Both a TeNT producing seed strain (strain A) and a non-TeNT producing strain, originating from strain G761 which lacks the plasmid that harbors the gene that encodes for the protein [28], were grown at Sanofi-Pasteur under standard growth condition by using small-scale batch fermentation [9]. All culturing occurred in duplicate. Small aliquots of the bacterial medium were taken at different time points (S1= 41 h post inoculation; S2= 65 h post inoculation; S3= 89 h post inoculation; S4= 113 h post inoculation; S5= 136 h post inoculation, S6= 142 h post inoculation, S7= 143 h post inoculation where the bacterial growth was stopped by cooling at 10 °C and by the addition of salt (final concentration of 2% (w/v) NaCl and 1.3% (w/v) $NaHCO_3$) to improve cell lysis and this until time point S8= 160 h post inoculation.

The aliquot of the cell suspensions was first centrifuged (5 minutes at 4000 x *g*) and the supernatants was subsequently filtered through a 0,22 µm filter to make sure no remnant bacteria were present. Next, 110 µL of 10 x protease inhibitor mixture (Roche) was added to 1mL supernatants, followed by a quick mix by vortexing, storage, shipment, and subsequent storage at Sciensano at temperatures below -70°C.

2.1.2. Different stress conditions

Additionally, to the standard conditions, the TeNT producing strain was also subject to three different stress conditions, including lowering of the pH of the growth medium to pH 6.0, cessation of the agitation during culturing and an increase of the incubation temperature by 2°C, to 36°C. All culturing occurred in duplicate, and aliquots of the culturing medium were taken at eight different time points during production. A schematic overview of the different sampling conditions is given in Figure 5A. These samples underwent the same procedure as described in 2.1.1.

2.1.3. Determination of the optical density (OD) and flocculation test

Optical density (OD) reflects the growth monitoring of *Clostridium tetani*. OD values are obtained by measuring cell suspension at 600 nm. If required, cell suspension was diluted in 0.9 % NaCl.

The amount of TeNT present in the in the culture supernatant was determined by means of a flocculation test, conform the test described in the European Pharmacopoeia [29]. Since this assay is known not to be sensitive and requires a high volume of supernatant, the concentration measurement was not performed on the first 2 sampling points, S1 and S2.

## 2.2. Sample preparation for mass spectroscopic analysis

2.2.1. Stock Solutions and Reagents

A stock solution of 25 Lf/ml TeNT was generated with the extracellular medium from non-TeNT producing bacteria and purified TeNT.

Reagents: Acetonitrile and methanol were ULC-MS grade and purchased from Biosolve (Valkenswaard, the Netherlands). Water was obtained using a Milli-Q Gradient A10 system (Millipore, Billerica, MA, USA). Analytical grade formic acid was bought from Merck (Darmstadt, Germany) while ammonium bicarbonate (purity > 99 %) Dithiothreitol (> 98 % HPLC purity), iodoacetamide (purity > 99 %) and Tween-20 were obtained from Sigma-Aldrich (St. Louis, USA). Sequencing grade modified trypsin from Promega (Wisconsin, USA) is used as digestion enzyme. Sequencing grade modified trypsin was purchased from Promega (Madison, WI, USA). All reactions were performed in Eppendorf LoBind protein tubes.

2.2.2. *In Silico* Selection of Peptides

UniProt release 2016_11 was used to obtain the different available TeNT sequences. The database UniProt Knowledgebase (UniProtKB, Magrane and the UniProt consortium, 2011) was searched for the terms “tetanus neurotoxin”. Multiple sequence alignment was performed using Clustal O v. 1.2.2 [30]. Next, the different proteins were subjected to an *in silico* digestion with trypsin by using Peptide cutter (Expasy). Only those peptides that were present in all protein sequences, contained at least six amino acids to a maximum of 20 amino acids, and for which the lowest digestion probability exceeded 80%, were included. To determine the specificity of the candidate peptides, the sequences were queried with BLASTP + 2.4.0 in NCBI using the default parameters. Only those peptides that were specific for TeNT were retained. Proteotypic peptides containing amino acids susceptible to side chain reactions were omitted. Additionally, the very hydrophobic and very hydrophilic peptides were rejected since they can either be problematic due to solubility issues or result in LC retention time instability respectively [31]. Of the remaining peptides, those with internal proline or glycine were favored as they are known to generate intense fragments, resulting in a panel of 8 candidate peptides (see supplemental Table 1).

2.2.3. Internal Standards (IS and ISd) and external calibrant

Peptide standards were custom synthesized to a purity level of ≥ 98 % (Thermo Fisher Scientific, Waltham, MA, USA). Stable heavy isotope labeled versions of the two selected peptides, belonging to the light chain (Lc) or the heavy chain (Hc) of TeNT respectively were also generated by Thermo Fisher scientific (purity GLD(I*)YYK = 98.48 %; purity VGYNAPG(I*)PLYK = 99.64 %). Stock solutions of 40 ng/mL were made by dissolving the IS in Milli-Q water and were stored at -20°C. In addition, a chimeric peptide, composed of two altered sequences (3 amino acid substitutions) of the above mentioned peptides (VAYQAPGIPLYKGLEIYYK; purity 98%) was used to determine the digestion efficiency and act as positive control, termed ISd. Standard stock solutions of ISd at 40 ng/mL were made by dissolving the chimera in Milli-Q water and stored at -20°C.

For each TeNT quantification experiment, a calibration curve was generated using the stock solution of TeNT and diluted to the desired concentrations (0.25; 1.25; 6.25; 18.75 and 25 Lf/mL) with medium from non-TeNT producing bacteria (external calibrant). Final concentrations of both IS (0.85 ng/mL) and ISd (40 ng/mL) were then added to the either the calibration curve or the samples that were to be analyzed.

2.2.4. Sample clean-up and tryptic digest

Prior to the enzymatic digestion, the samples underwent a simple sample clean up step to remove interfering substances. Briefly, 50 µL of each aliquot was transferred to a 1.5 mL Eppendorf LoBind tube (Eppendorf, Hamburg, Germany) and 450 µL of cold methanol (stored at -20°C) was added, followed by vortexing and incubation for 20 min at -20°C. Next, the tubes were centrifuged at 20 000 x *g* for 5 minutes and 450 µL of supernatant was removed. The pellet was washed once with methanol prior to the drying of the samples at 60°C and subsequent incubation at -20°C for at least two hours. The sample was first reconstituted with 50 µL of PBS + 0.1 % (v/v) Tween-20 and underwent the addition of the control peptides (IS and ISd), reduction with dithiothreitol (final concentration of 15 mM, incubation for 20 min at 99°C) and alkylation of the cysteines by iodoacetamide (final concentration of 20 mM, incubation for 30 min in the dark at room temperature). Next, the mixture was diluted with 0.1 M ammonium bicarbonate pH 8.0 to a final volume of 470 µL containing 5 µg of trypsin per sample. The digest was incubated overnight at 37°C and the reaction was stopped by adding 10 µL of a 10 % (v/v) formic acid solution, followed by a vacuum concentration step (5 h at 45°C). The pellet was first incubated at -20°C for at least two hours prior to reconstitution with 500 µL of 1 % (v/v) formic acid solution. After digestion the peptides were centrifuged at 20 000 x *g* for 5 minutes and the supernatant was transferred to injection vials.

### 2.3. Instrumental conditions LC-MS/MS

The UPLC-MS/MS system consisted of an Acquity UPLC system in line with a Xevo TQ-S triple quadrupole mass spectrometer (Waters, Milford, MA, USA). The chromatographic separation of 10 µL of digested and labelled peptides was performed with a mobile phase consisting of 0.1 % formic acid in water (A) and 0.1 % (v/v) formic acid acetonitrile (B) at a total flow rate of 0.3 ml/min while the column temperature was kept at 45°C. Peptide separation was performed with an ACQUITY UPLC CSH C18 Column (2.1 x 150 mm, 1.7-µm particle size). The elution method employed is an optimized gradient starting at 3% B to 70% B in 8 min followed by a column wash and equilibration steps. Briefly, a 1 minute isocratic elution with 3 % is followed by, a linear increase to 14% over one min, a linear increase to 20 % in 2 min, an increase to 70 % in 4 min, followed by a 2 min wash step at 99 % B and a 2min re-equilibration step at 3 % B. The samples were ionized by means of an electrospray probe in positive mode operating at a cone voltage of 30 V for the IS and at 45 V for the ISd. The capillary voltage was set at 2.25 kV. The selected reaction monitoring (SRM) method acquisition method was constructed using the following source parameters: 2.25 kV spray voltage, 150°C source voltage, 500°C desolvation temperature, 150 L/h cone gas flow, 1000 L/h desolvation gas flow, 0.15 mL/min collision gas flow, and 7 Bar nebulizer gas flow. The collision energies were optimized for each selected peptide, IS and ISd. The selected quantification and confirmation fragments with their respective collision energies are listed in Table 1. The dwell time for both transitions (except for the ISd where one single transition served as quantifier and qualifier) was set in order to acquire at least 12 data points.

## 2.4 Validation of the method

Validation of the TeNT quantitation method was performed based on ICH guidelines [32] and the total error approach [33, 34].

2.4.1. Selectivity, specificity, LOD, LLOQ, ULOQ, linearity of the calibration line and matrix effect

Briefly, the selectivity and specificity of the method was assessed. In addition, the limit of detection (LOD), the lower limit of quantification (LLOQ), and the upper limit of quantification (ULOQ) were determined by means of the output of the external calibrant. Additionally, linearity was assessed for TeNT concentrations ranging from 0.25 to 25 Lf/ml by applying least squares regression analysis. Adequate linearity was achieved when the regression coefficient (r) ≥ 0.99 and non-linearity was tested with a Mandel's fitting test as described in [35,36]. Potential matrix effects were evaluated by comparing the obtained amount of TeNT at the LLOQ and ULOQ at 3 different time points (described in 2.1.1.). All injections were performed in quadruplicate.

2.4.2 Trueness, accuracy, precision, and uncertainty by means of the total error approach

This approach estimates the "total error" (TE) by combining the systemic error (bias) and the random error (intermediate precision) to determine the difference between the observed result and the true value. In other words, the highest error of an analytical method can be estimated. The TE (%) is calculated as bias (%) + 1.65 × intermediate precision (%). The factor 1.65 implies that 95 % of the results will fall within the TE limit, given a Gaussian distribution.

For this experiment, TeNT-spiked blank samples (medium of non-TeNT producing bacteria at the S4 time point) were made daily in triplicate at five concentration levels (0.25; 1.25; 6.25;

18.75 and 25 Lf/mL) and analyzed for at least six consecutive days. Also, each day a calibration curve was generated for six concentrations (0.25; 1.25; 6.25; 12.5; 18.75 and 25 Lf/mL). The corresponding concentrations were calculated from the measured peak areas using the calibration curve. These calculated concentrations (i.e., the 5 concentration levels) were used to determine, for this concentration range, the applicability of linear relationship between the measured and theoretical concentrations, the trueness, precision (repeatability and intermediated precision) and accuracy, by means of a validated Excel spreadsheet [36-38]. The β-expectation tolerance limits, calculated at each concentration level, can be used as a predictive tool such that that 95 % of future results using this analytical method are expected to fall within the predefined acceptance limits [−λ; λ]. For the current method, the acceptance limits were, in agreement with the industrial partners, predefined to be within a target range of ± 30 %, with the exception of the LLOQ, where a target range of ± 40 % was deemed to be acceptable.

## RESULTS AND DISCUSSION

### 3.1 Choice of quantitative targeted LC-MS design

There are currently several LC-MS/MS approaches described in the literature to quantify the amount of protein in the sample [40]. The indirect method, via quantification of the peptides after digestion by estimating the concentration of this peptide based on the LC-MS/MS response of the internal standard, which is often a stable isotope labeled (SIL) surrogate peptide. This methodology compensates for the variation resulting from the mass spectrometry instrumentation itself but does not compensate for the variability introduced in the digestion step as this will generate most of the variation in the outcome [39]. To this end, extended peptides or chimeric peptides, consisting of the proteotypic SIL peptides, are sometimes used. However, even these peptides cannot mimic an entire protein. Therefore, recently SIL versions of the protein of interest gained popularity since they will enable the tracking of the protein throughout the entire analytical procedure [40]. SIL proteins are generally added at the start of the sample extraction and can account for almost all variation ranging from sample preparation, enzymatic digestion, pre-analytical treatments as well as the mass spectrometric ionization. However, a huge drawback for the use of SIL proteins as internal calibrants for protein quantification is their high cost of production. Alternatively, hybrid methodologies, using an external calibrant with internal standards can be used, generally meaning that the purified version of the protein of interest is used as calibrant when present in an analogous matrix which also contains SIL peptides [41-46]. This latter, more economically favorable methodology, requires the generation of proteotypic SIL peptides, the availability of reference protein and the availability of a very similar matrix. The latter is generated for our experimental set-up by the use of the growth medium of the non-TeNT producing *Clostridium tetani* strain whereby an external standard corresponding to a known amount of purified TeNT is run in parallel to the cell culture analyte samples, using the same SIL peptide standards and identical LC-MS/MS workflow (see Figure 1).

### 3.2. Development of the SIL internal standards, external calibrant and sample preparation

Prior to selection and procurement of the SIL peptide standards, a subset of the 8 specific tryptic peptides selected according to the *in silico* criteria (Section 2.2.2) were synthesized without use of heavy labels and screened using MRM (see supplemental Table 1). The most abundantly detected peptides for the light chain (GLDIYYK) and heavy chain (VGYNAPGIPLYK) respectively were retained. Moreover, we confirmed that these peptides were indeed present in 3 independent tryptic digest experiments of the purified TeNT that were performed according to the previously described settings [47]. Accordingly, heavy isotope labeled versions of those peptides, GLD(I*)YYK and VGYNAPG(I*)PLYK, were synthesized and used as the internal standards (IS). As mentioned above, SIL peptides cannot account for variability in the LC–MS/MS analysis introduced by upstream steps such as sample preparation and digestion [48, 49]. Therefore, for each quantification experiment of TeNT, a calibration curve was generated from the stock solution of purified TeNT of known content and diluted to the desired concentrations (0.25, 1.25, 6.25, 18.75, and 25 Lf/mL) with medium from non-TeNT producing bacteria (see Figure 1). As an additional internal control specifically to monitor the digestion (ISd), we also included a chimeric peptide consisting of a combination of the chosen TeNT peptides but with 3 amino acid substitutions so the peptides obtained from the chimeric synthetic peptide could be distinguished from the ones originating from TeNT (VAYQAPGIPLYKGLEIYYK). This chimeric peptide was used mainly during the digestion optimization process.

As a direct dilute-and-shoot experiment might not be recommended for the quantification of TeNT in complex matrices such as bacterial media, the use of an easy, low-cost and mass spectroscopy compatible sample clean-up by protein precipitation was performed and assessed for recovery by bicinchoninic acid (BCA) assay on the starting solution, pellet and dried supernatants (data not shown). To select the optimal conditions for sample precipitation, one volume of TeNT in cell culture matrix was incubated with 9 volumes of several different classical solvent systems that were each stored at -20°C prior to mixing (methanol, acetonitrile, methanol: acetonitrile (50:50), and trichloroacetic acid/acetone). Optimal results were obtained with 100 % methanol and incubation at -20°C for at least 2 hours. The BCA results were confirmed by LC-MS/MS experiments where the recovery of the native peptides was compared between the precipitation conditions. Since we could not exclude the possibility that some fraction of the sample might not precipitate, the external calibration standard was subjected to the same clean-up and tryptic digestion protocol (see Methods section). In order to account for the impact of multiple protein impurities expected to be in the samples, the amount of trypsin used for the digestion was optimized using a fixed amount of TeNT (25 Lf/mL) that was spiked in non-TeNT medium. It was found that the addition of 5 µg of trypsin generated stable and reproducible results and moreover, no signal increase was observed for higher amounts of trypsin.

### 3.3. Development and optimization of the LC-MS/MS method

The UHPLC conditions were developed and optimized using a Waters Acquity CSH C18 Column (2.1 × 150 mm; 1.7 µm particle size). An optimized UHPLC gradient, taking 12 minutes in total, including cleaning and equilibration steps generated different retention times among the cluster of natural peptides, co-eluting IS peptides and the ISd fragment, and the cluster containing the ISd peptide. The optimal selective reaction monitoring (SRM) conditions were determined for each peptide separately by direct infusion and verified by testing different SRM conditions for samples that consisted of the different peptides present in the resuspended and digested protein pellet obtained from the non-TeNT containing medium.

### 3.4. Validation of the targeted LC-MS/MS method

3.4.1. Specificity, selectivity, LOD, LLOQ, ULOQ, linearity of the calibration line and matrix effect

In order to develop an appropriate methodology to quantify the amount of TeNT during the production process, it is essential that the method is able to correctly identify and distinguish the selected peptides of interest from the matrix ingredients and from each other. This can be achieved by combining the two separation mechanisms of LC and MS/MS allowing the analysis of complex mixtures. The specificity and the selectivity of the utilized targeted LC-MS/MS method is illustrated in Figure 2 (quantifier transitions) and supplemental Figure 1 (qualifier transitions). This must be demonstrated for the sample concentration corresponding to the limit of detection (LOD) at which the least responsive peptide (the Lc peptide: GLDIYYK) can be correctly identified in all TeNT containing samples with a signal to noise ratio greater than 3.3 and is absent in the samples originating from non-TeNT producing culture. In addition, the lower limit of quantification (LLOQ) was determined as the lowest concentration where precision and accuracy could be demonstrated (see 3.4.2). Furthermore, the LLOQ must have a signal to noise ratio $\geq 10$ and was also assigned as the lowest point of the calibration curve. The upper limit of quantification (ULOQ) was determined as the highest sample concentration where precision and accuracy could be demonstrated (see 3.4.2).

The LOD was determined to be equivalent to 0.025 Lf/mL TeNT, the LLOQ was 0.25 Lf/mL and the ULOQ was 25 Lf/mL. Moreover, the relationship between the responses was shown to be linear for all 6 concentration levels assessed within the concentration range of 0.25-25 Lf/mL. The linearity of this six point calibration curve was assessed via least square regression. All $R^2$ values were above 0.98. In order to verify the possibility of a quadratic relationship, a Mandel's fitting test was performed. A linear model was preferred as all the calculated F-values were lower than the critical F-values. This quantification range was judged to be acceptable since the Lf assay has an LLOQ for TeNT close to the ULOQ of the LC-MS/MS method. Moreover, the developed method is almost 10-fold more sensitive than a previously described LC-MS/MS based TeNT detection method [18]. When compared to a specific quantitative ELISA test for TeNT quantitation, the proposed LC-MS/MS method appears to be approximately 25-fold less sensitive [20]. However, it should be mentioned that the LODs obtained for the ELISA methods, are obtained with either the purified TeNT or purified toxoid, whether or not in the presence of adjuvants, and thus may not reflect the real-life conditions when working with the detection and quantification of a protein in complex bacterial medium. Additionally, potential matrix effects should be evaluated as the presence of salts or other peptides could result in assay variability. As this method was developed to assess the amount of TeNT present in bacterial media at various stages of fermentation, we evaluated the amount of TeNT present in the medium at three different time points (see 2.4.1). All injections were performed in quadruplicate and subjected to a t-test, which demonstrated no matrix effect. However, the use of bacterial medium prior to the inoculation with bacteria demonstrated a significant effect (data not shown). Therefore, the current method can be utilized for the quantification of TeNT starting from 19 h post inoculation.

3.4.2. Validation of the quantification by means of the total error approach

The validation of the quantification method was performed according to the general ICH guideline and by using the total error approach [32-34]. Unfortunately, there are no predefined validation acceptance limits that are directly applicable on the quantification of TeNT during vaccine manufacturing. However, based on matrix complexity and the overall purpose of the assay, the guideline ICH M10 for bioanalytical assays could be applied [50]. Three different quantification strategies were assessed based on the use of either the Lc or the Hc peptide as quantifier or as the mean value obtained when both the Lc and the Hc peptide served as quantifier. The generated results are summarized in Table 2 and Figure 3.

3.4.2.1. Linearity of the results

Linearity is the ability of an analytical procedure to provide test results which are proportional to the concentration. This linear correlation between the theoretical and measured concentrations (for the predefined concentration range) is expressed by the $R^2$ which must be ≥ 0.99. The linearity of the method is acceptable as $R^2$ values are above 0.999 (see Table 2).

3.4.2.2. Trueness

The trueness of an analytical procedure expresses the closeness of agreement between the average value obtained from repeated measurements and the theoretical "true" value (termed "accuracy" according to ICH guidelines). This is a measure of the systematic error of the method and is expressed in terms of relative bias. According to ICH M10, the trueness is deemed acceptable if relative bias between the theoretical and measured values for a given concentration level is within ± 15 %, except at the LLOQ, where it must be within ± 20 %. From Table 2 it can be concluded that the trueness for all components is acceptable since the relative bias is in all cases ≤ 10.46 %.

3.4.2.3. Precision

The precision of an analytical method expresses the closeness of agreement between the values obtained from repeated measurements. This was investigated at two levels: (i) the repeatability, which gives the precision under the same operating conditions over a short time interval (in-run), and (ii) the intermediate precision, also termed reproducibility, which expresses the intra-laboratories variations (between-run), assessed on different days. The precision is a measure of the relative error of the method and is expressed in terms of relative standard deviation (RSD). The precision is deemed acceptable if the intermediate precision and the repeatability are below 15 % for all concentration levels, except at the LLOQ where 20 % is allowed. From Table 2 it can be concluded that the method performance is acceptable since all RSD values were ≤ 12.63 %).

3.4.2.4. Accuracy

Accuracy takes into account the total error (TE) of the test results and is represented by the ß-expectation tolerance limits. These limits, calculated at four different concentration levels, can be used as a predictive tool such that 95 % of future results using the analytical method would be expected to fall within the predefined acceptance limits [−λ; λ]. For the current method, we assigned acceptance criteria for accuracy such that the β-expectation tolerance limits for the concentration levels studied are ± 30 % for all concentration levels, except at the LLOQ where the limits are ± 40 %, in accordance with ICH M10 guidance. For all concentration levels except at the LLOQ for the Hc peptide (value of ± 30.99 %), tolerance limits were lower than ± 30 % target value. As the predefined acceptance criteria were met, the method was judged fit for its intended purpose in terms of accuracy.

3.4.2.5. Expanded uncertainty

The expanded uncertainty is calculated as a 95 % confidence interval around the reportable results in which the true value can be found. As shown in Table 2 the maximum observed value for relative expanded uncertainty, calculated as the ratio of the expanded uncertainty at a 95 % confidence level and the respective concentration level, corresponds to 26 %, which was deemed as a scientifically reasonable margin for this type of analysis.

### 3.5. Assay format and system suitability test for routine analysis

The next step was to apply the validated assay to a real-life TeNT fermentation process. This includes the presence of at least one technical replicate of the sample preparation. Moreover, as both the use of one quantifier peptide to quantify the amount of TeNT, and the use of the mean of values obtained for both peptides were in scope of the method validation, it was decided to choose the latter methodology. We could also incorporate two additional validity criteria, which is highly desirable for an in-process control assay (see Figure 4). If the samples are prepared in duplicate and subjected to quantification by means of each peptide separately, we can incorporate an additional validity criterion when determining the mean of both measurements. Indeed, based on the data generated during the validation, the maximum relative standard deviation (% RSD), of the amount of TeNT quantified during one single injection, either by utilizing the Lc peptide as quantifying peptide or alternatively by utilizing the Hc peptide as quantifying peptide corresponded to 18.44 % at the LLOQ (see supplemental table 2). Moreover, the overall mean % RSD and the median % RSD values were close to each other, indicating that the method is robust. In addition to this in-sample validity criterion, a second between-sample validity criterion was proposed. The second criterion, which is based on the relative expanded uncertainty obtained for the 3$^{rd}$ strategy (see 3.4.2.5.) was set at 25 %. If these assay validity criteria were not met for the technical duplicates, the experiment was deemed invalid and had to be repeated. A schematic overview of the experimental strategy for a real-life routine experiment is depicted in Figure 4. This experiment was performed on each of the fermentation samples (see 3.6).

### 3.6. Application of the method during a real-life experiment: assessment of the effect of different stress conditions on the quantity of TeNT being produced

TeNT is produced by fermentation using a complex growth medium at pH 7.2, under continuous nitrogen flow into the medium, a growth temperature of 34°C, and under constant agitation. At the end of the fermentation, approximately 142 h post inoculation, the toxin is harvested from the culture supernatant, inactivated with formaldehyde to produce the toxoid, which is then purified and combined with other antigens for use in multivalent vaccines [11]. Past reports have shown that growth on a more acidic medium resulted in lower to no TeNT production at the time of harvesting. The same study also demonstrated that an increase in growth temperature to 37°C resulted in a slight decrease in the amount of TeNT present in the medium [51]. In addition to these physical parameters, growth medium composition plays a critical role in TeNT production [11, 22, 52]. However, attempts to rationally design medium have resulted in reduced TeNT production [11]. Therefore, in order to mimic real life stress factors that could occur during the industrial production process, we focused on the physical parameters such as an augmentation of the growth temperature (from 34°C to 36°C), lowering of the pH (to pH 6.0) and a cessation of constant agitation. TeNT culturing was performed in duplicate for the abovementioned stress conditions and the normal growth condition. Both the optical density of the culture (to monitor bacterial growth) and the cell viability were monitored in addition to determination of the amount of TeNT (Figure 5B). The latter was determined by both the traditional flocculation assay (Figure 5C) and the developed LC-MS/MS quantitation method (Figure 5D and Supplemental Table 3).

The normal TeNT fermentation process is characterized by a bacterial growth curve with maximum bacterial cell density achieved around 90 h post inoculation (Figure 5B, blue lines). From that point, the bacteria begin to undergo lysis, resulting in the release of intracellular TeNT into the medium. A sharp increase in extracellular TeNT levels from 90 h to 140 h is followed by a slower increase and stabilization or even a slight decrease for the subsequent time points. The stabilization phase is due to the fact that there are no longer TeNT producing bacteria and the slight decrease could be explained by the presence of intracellular proteases which co- released into the medium. This general pattern of TeNT expression is evidences by both the flocculation test and the targeted LC-MS/MS assay (Figure 5C and D, blue lines).

However, the absolute values differ by a factor of 1.5 to 2 at the higher levels of TeNT in the medium. This may be explained by the fact that the Lf test is only semi-quantitative [17, 20], and is normally performed on the purified toxoid rather than TeNT. Moreover, there may be interference with the flocculation assay from other *C. tetani* proteins or other components of the bacterial expression medium [17]. This later hypothesis can explain the discrepancy in the amount of TeNT found during the two last time points, where active bacterial cell lysis was induced by the addition of salt, between the Lf assay and the developed methodology (see Figure 5C and 5D, blue lines).

When comparing the standard conditions to those in growth medium with lower pH, it is obvious that the lowering of the pH had a detrimental effect on the amount of TeNT produced (Figure 5C and D, green lines). This was likely due to the lack of bacterial growth observed under low pH conditions (Figure 5B, green lines). This is consistent with what is known in the literature [50]. An increase in temperature of the culture (from 34˚C to 36˚C) resulted in an initial augmentation of TeNT expression compared to standard conditions, but then a plateau and drop-off at the later harvesting points (Figure 5C and D, red lines). The early augmentation is likely due to increased lysis of bacteria earlier in the fermentation compared to the normal growth conditions (Figure 5B, red lines) and the subsequent release of the previously intracellular TeNT into the medium [53]. Unfortunately, this phenomenon at the intermediate time points could not be verified by means of the flocculation test due to the much higher limit of detection of that assay. From 120 hours post inoculation onwards, a decrease in the amount of TeNT present in the medium was observed by both LC-MS/MS and flocculation assay and is likely due to proteases that would also be expected to be released during bacterial lysis. The condition with no agitation of the growth medium, on the other hand, showed similar bacterial growth profiles to the standard conditions (Figure 5B, orange lines) as well as similar TeNT expression, as assessed by both the flocculation assay and by the developed LC-MS method (Figure 5 C and D, orange lines).

## CONCLUSION

A targeted LC-MS/MS method was developed and validated for the purpose of quantitation of TeNT expressed by *Clostridium tetani* during the production process to generate the tetanus toxoid vaccine antigen. This method is able to detect TeNT present in the bacterial growth medium with a detection limit of 0.025 Lf/mL and is able to accurately quantify TeNT in the same complex matrix with a LLOQ of 0.25 Lf/mL. The developed method is almost 10-fold more sensitive than a previously described LC-MS/MS based TeNT detection method [18]. This latter method was sufficient only for qualitative identification of TeNT. Compared to a specific quantitative ELISA test for TeNT quantitation, our LC-MS/MS method appears to be approximately 25-fold less sensitive [20]. However, it should be mentioned that the LODs reported for the ELISA methods were obtained with either the purified TeNT or purified toxoid, and thus may not reflect real-life conditions for detection and quantification of a protein in complex bacterial medium.

The quantitative LC-MS/MS method was successfully applied to monitor the amount of TeNT produced during the multi-day production process, an important complementary method to optical density, which only monitors bacterial growth and not specific TeNT expression. Our methodology can also be used to support the optimization of culture process or to assess the impact of changes in culturing conditions. Furthermore, the quantification of TeNT in combination with transcriptomic data, under both standard growth conditions and stressed conditions could give valuable information about TeNT gene regulation and the biology of *C. tetani* cell culture [22].

## FIGURES AND TABLES

**Figure 1.** Scheme illustrating the experimental set-up, described in 3.1, for quantitation of TeNT

**Figure 2**. The LC-MS/MS chromatograms obtained for the quantifier transitions of the Lc (A) and Hc peptides (B), their SIL versions (termed IS), for the medium of non-TeNT producing bacteria after trypsin digestion and the medium of non-TeNT producing bacteria spiked with 0.25 Lf/mL TeNT prior to the digestion step and spiked with IS. The boxes indicate the presence of the Lc and Hc peptides while the arrows indicate the internal controls.

**Figure 3.** Accuracy profiles with β-expectation interval (solid red line), relative bias (solid blue line), 95 % β-expectation tolerance limits (dashed green line); back-calculated concentrations of reference standards (dots) and ± 30 % acceptance limits (red line). These accuracy profiles were established for the back calculation strategy 1, where the Lc peptide is taken as quantifier peptide (A), strategy 2, were the Hc peptide is taken as quantifier (B) and strategy 3, where each peptide separately serves a quantifier and where the mean of the obtained values is reported (C).

**Figure 4.** Scheme illustrating the experimental set-up for a real-life routine quantitation of TeNT. For each routine sample analysis, the sample will be split into two aliquots (A1 and A2), both subjected to separate samples clean–up, tryptic digest and injection. The mean of the back calculated values, based on the area obtained for the two peptides, will be averaged. The % relative standard deviation for this averaging should be lower than 20 %. Next, also the value obtained for both aliquots is averaged to obtain one final end value. Also here, an additional validation criterion was introduced stating that the relative standard deviation for this averaging

should be lower than 25 %. If both validity criteria are not met, the obtained values are not accepted and the analysis has to be redone.

**Figure 5.** The effect of different stress conditions (A) on bacterial growth (B) and the concentration of TeNT present in the medium as assessed by the flocculation assay (C) or by targeted LC-MS/MS (D). The values obtained for the standard conditions are depicted in blue, while the pH stress are represented in green, the heat stress conditions in red and the no agitation conditions in orange. All experiments were performed in duplicate (biological replicate). The arrow represents the time point where salt was added to the culture, to induce additional cell lysis.

**Table 1**. Summary of the selective reaction monitoring characteristics of the selected peptides, also including their sequences, their location on the protein and their retention time. IS= SIL labeled internal standard, ISd= internal standard for the digestion. The asterix indicates the amino acid which is labeled.

**Table 2.** Trueness, precision, accuracy and relative expanded uncertainty of the chromatographic methodology for the three different back calculation strategies (RSD: relative standard deviation).

**Supplemental Figure 1.** The LC-MS/MS chromatograms obtained for the qualifier transitions of the Lc and Hc peptides, the SIL version of these peptides (IS), the tryptic digest control chimera (ISd) and the ISd fragment, obtained after digestion with trypsin of the medium of non-TeNT producing bacteria (A) of the medium of non-TeNT producing bacteria that were spiked with IS and ISd (B) and the medium of non-TeNT producing bacteria spiked with 0,25 Lf/mL TeNT, IS and ISd (C). The arrows indicate the internal controls. The asterix indicates the disappearance of the ISd chimera while the boxes demonstrate the presence of the Lc and Hc peptides appearing only after tryptic digest.

**Supplemental Table 1.** The sequences and characteristics of the panel of 8 remaining possible proteotypic peptides and the flow chart utilized to select the most intense ones.

**Supplemental Table 2.** The mean, median and maximum percentage relative standard deviation (%RSD) for the amount of TeNT quantified, during one single injection, either by utilizing the peptide GLDIYYK as quantifying peptide or alternatively by utilizing the peptide VGYNAPGIPLY as quantifying peptide (total number of injections per concentration, n=18).

**Supplemental Table 3.** The amount of TeNT present in the bacterial medium as determined by the targeted LC-MS/MS method (the values are expressed in Lf/mL). Different stress conditions were compared (see text for details on the stress conditions).

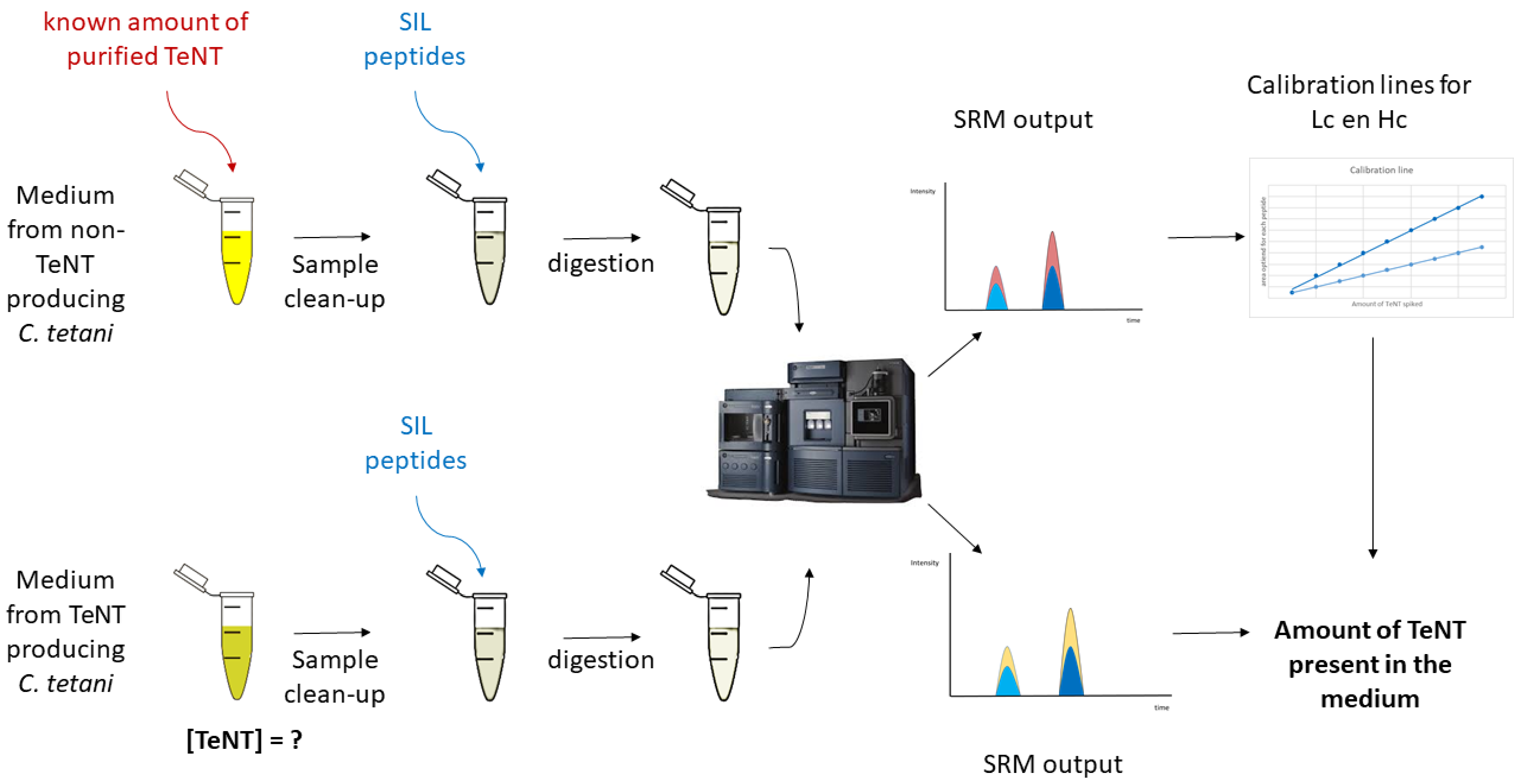
known amount of purified TeNT
SIL peptides
Medium from non-TeNT producing *C. tetani*
Sample clean-up
digestion
SRM output
Intensity
time
Calibration lines for Lc en Hc
Calibration line
Amount of TeNT spiked
SIL peptides
Medium from TeNT producing *C. tetani*
[TeNT] = ?
Sample clean-up
digestion
Intensity
time
SRM output
Amount of TeNT present in the medium

**Figure 1**

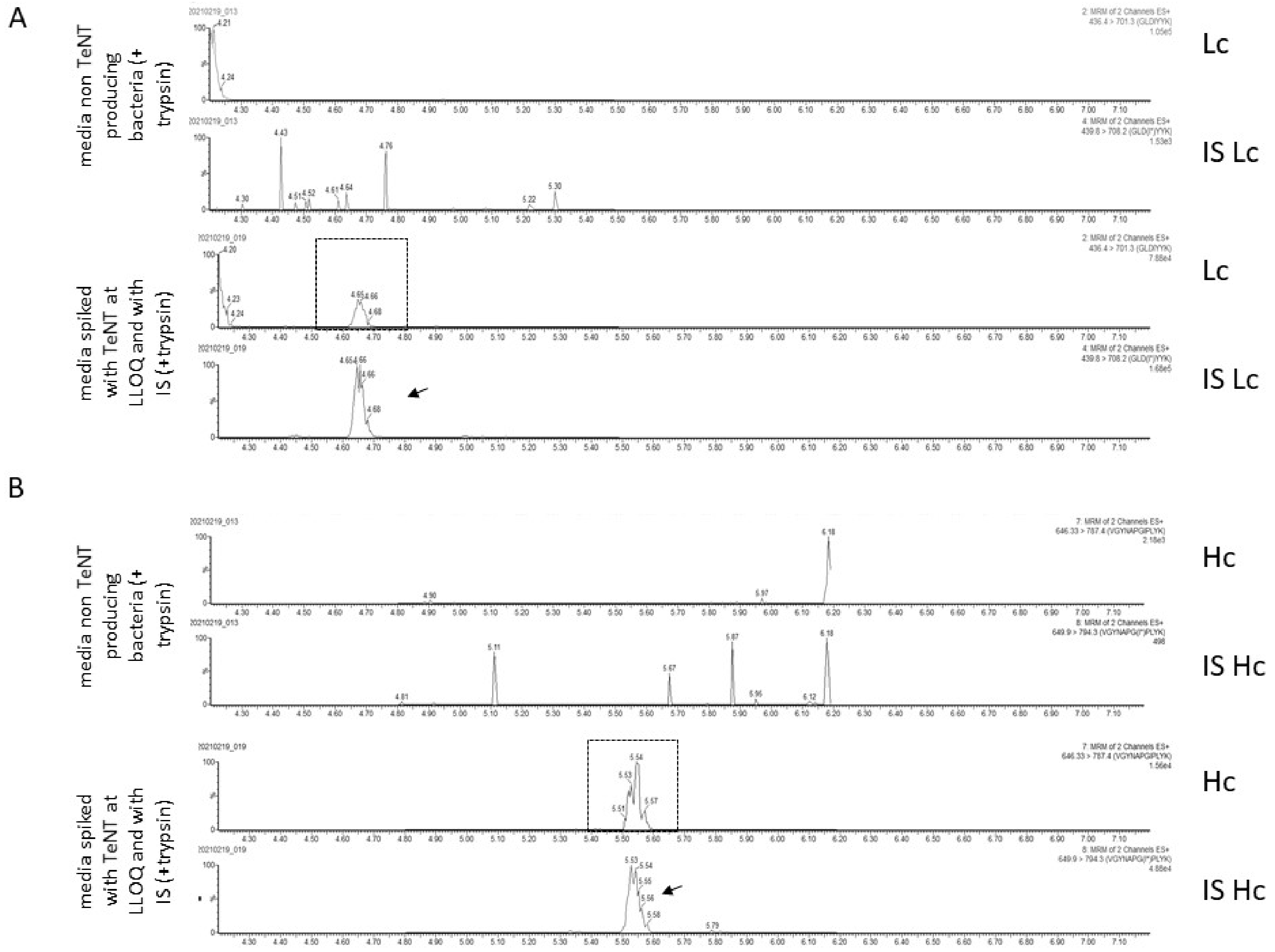
A
media non TeNT producing bacteria (+ trypsin)
media spiked with TeNT at LLOQ and with IS (+trypsin)
Lc
IS Lc
Lc
IS Lc
B
media non TeNT producing bacteria (+ trypsin)
media spiked with TeNT at LLOQ and with IS (+trypsin)
Hc
IS Hc
Hc
IS Hc

**Figure 2**

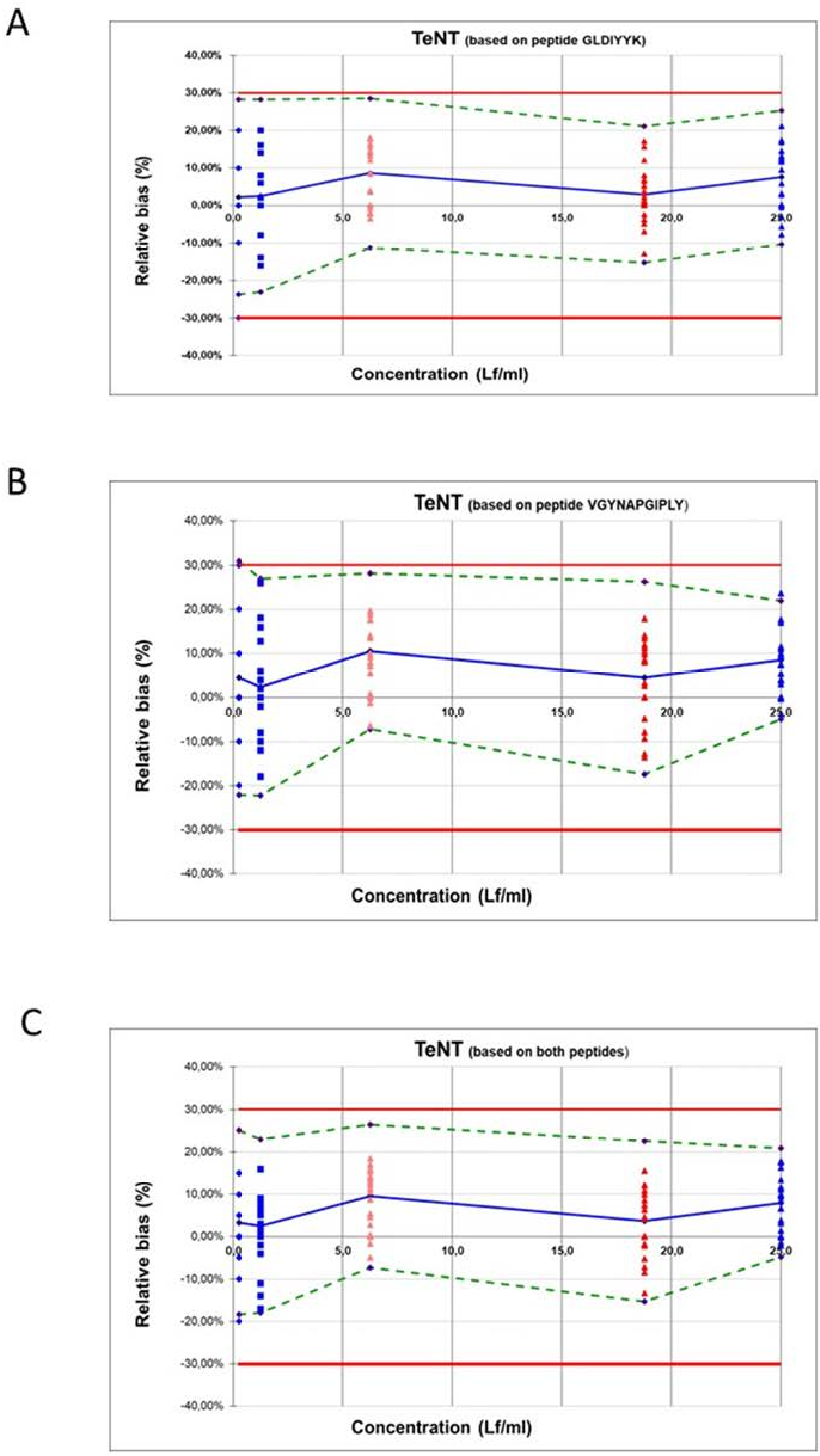

**Figure 3**

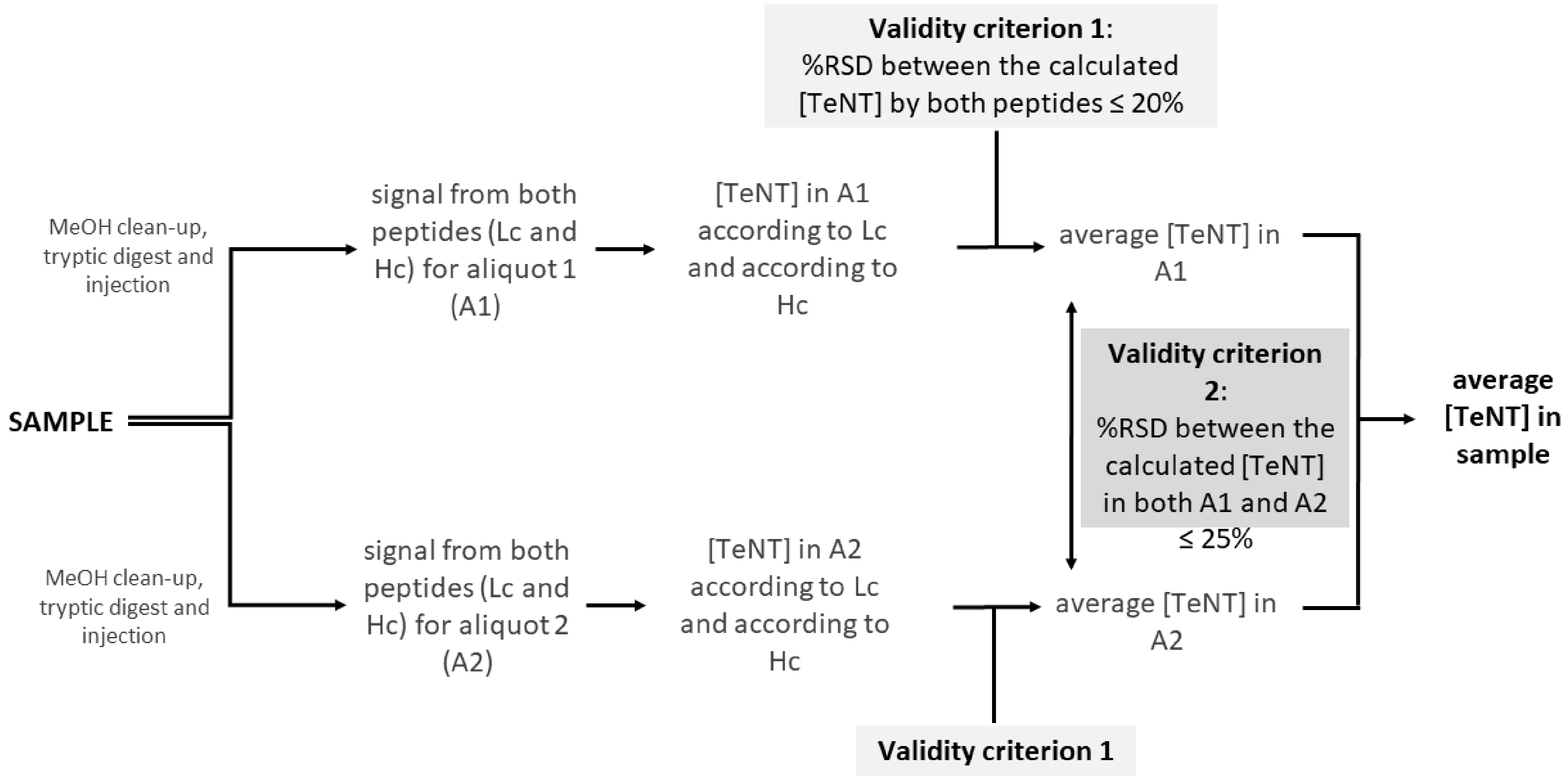
Validity criterion 1:
%RSD between the calculated [TeNT] by both peptides ≤ 20%
MeOH clean-up, tryptic digest and injection
signal from both peptides (Lc and Hc) for aliquot 1 (A1)
[TeNT] in A1 according to Lc and according to Hc
average [TeNT] in A1
Validity criterion 2:
%RSD between the calculated [TeNT] in both A1 and A2 ≤ 25%
SAMPLE
average [TeNT] in sample
MeOH clean-up, tryptic digest and injection
signal from both peptides (Lc and Hc) for aliquot 2 (A2)
[TeNT] in A2 according to Lc and according to Hc
average [TeNT] in A2
Validity criterion 1

**Figure 4**

A

| Abbreviation | Culture conditions |
|---|---|
| Batch 1 | Standard: 34°C, pH 7.3, 150 rpm stirring |
| Batch 2 | Standard: 34°C, pH 7.3, 150 rpm stirring |
| pH 6_1 | pH stress (pH 6.0) |
| pH 6_2 | pH stress (pH 6.0) |
| NOA_1 | no agitation |
| NOA_2 | no agitation |
| HEAT_1 | 36 °C |
| HEAT_2 | 36 °C |

B

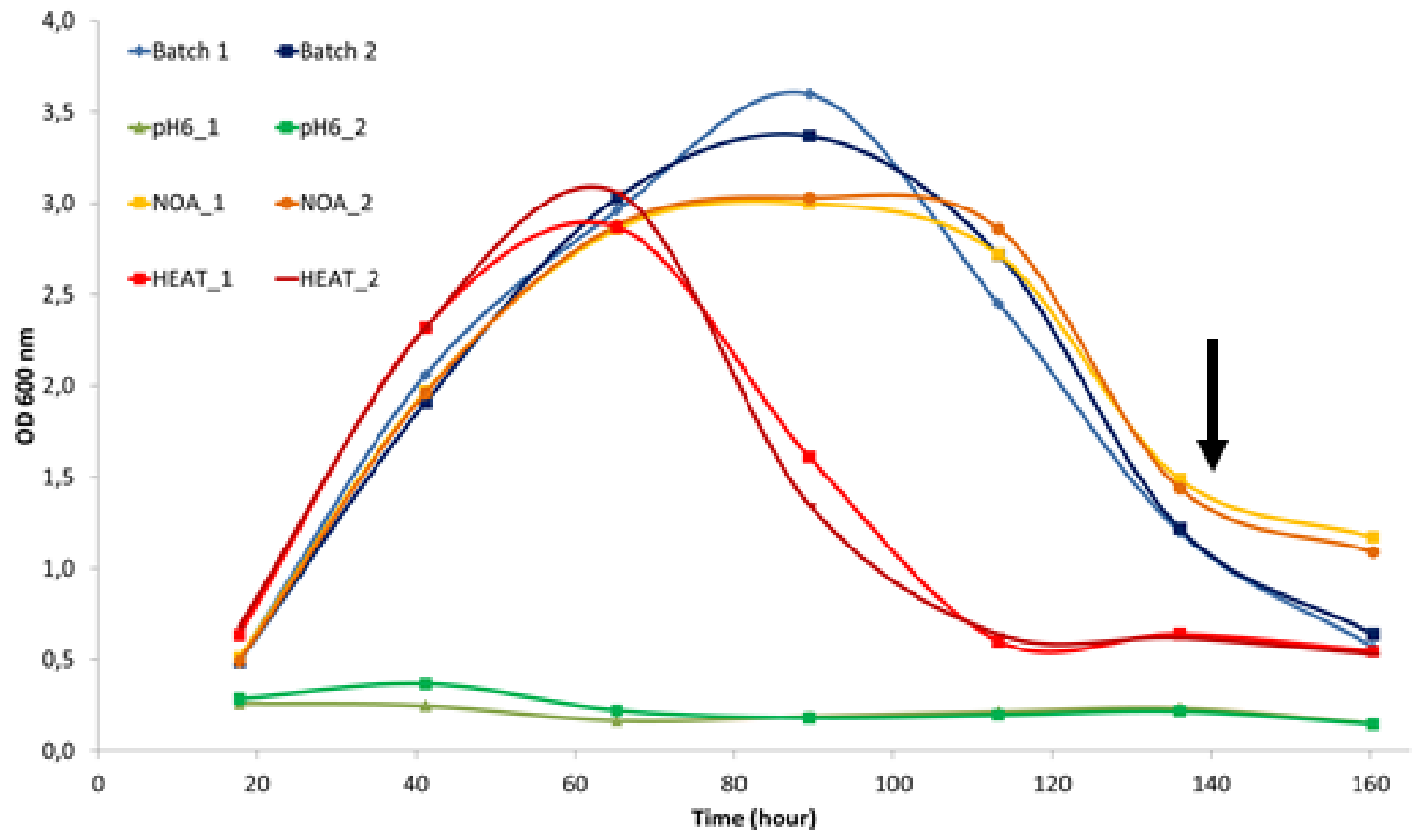

C

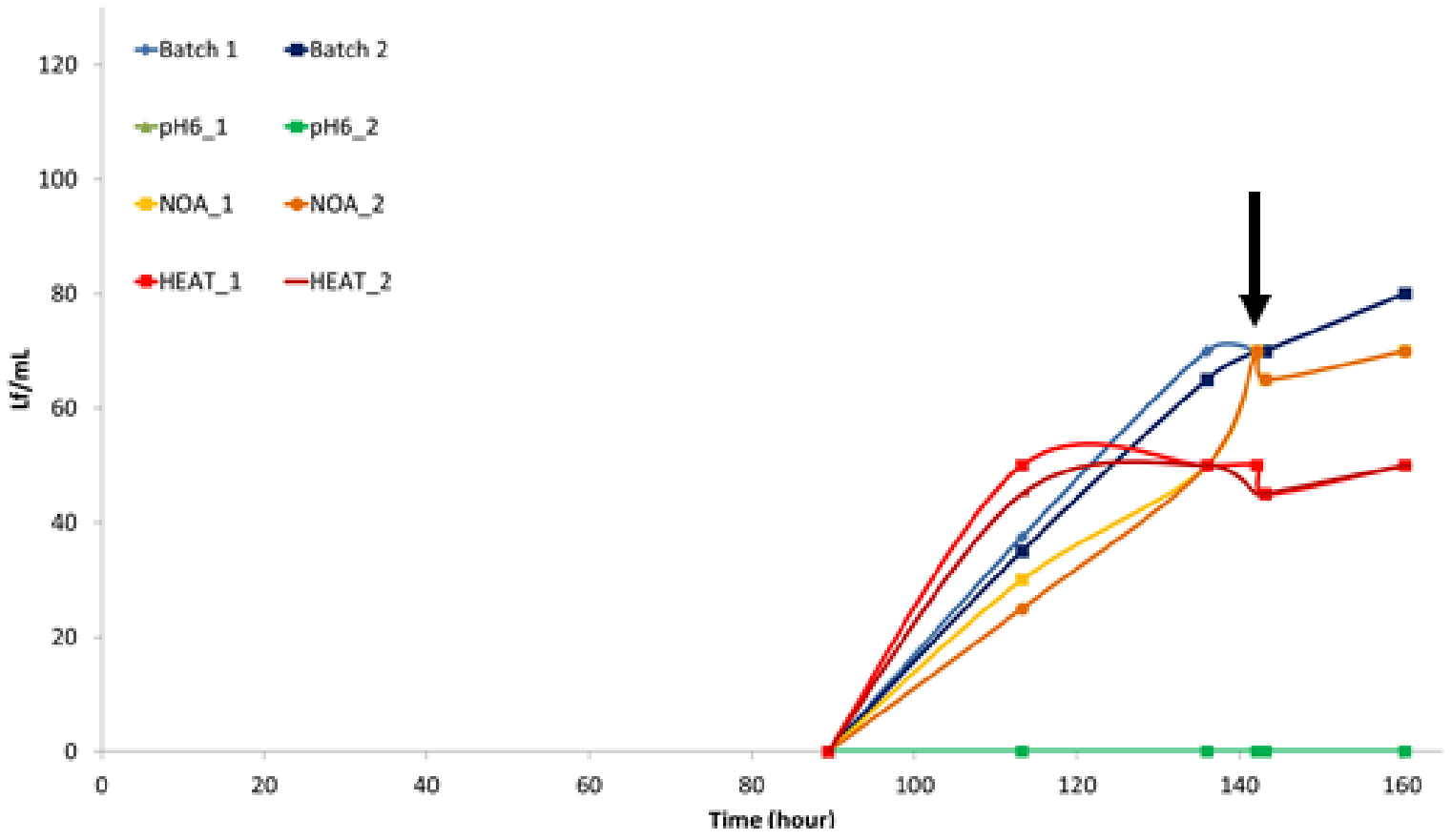

D

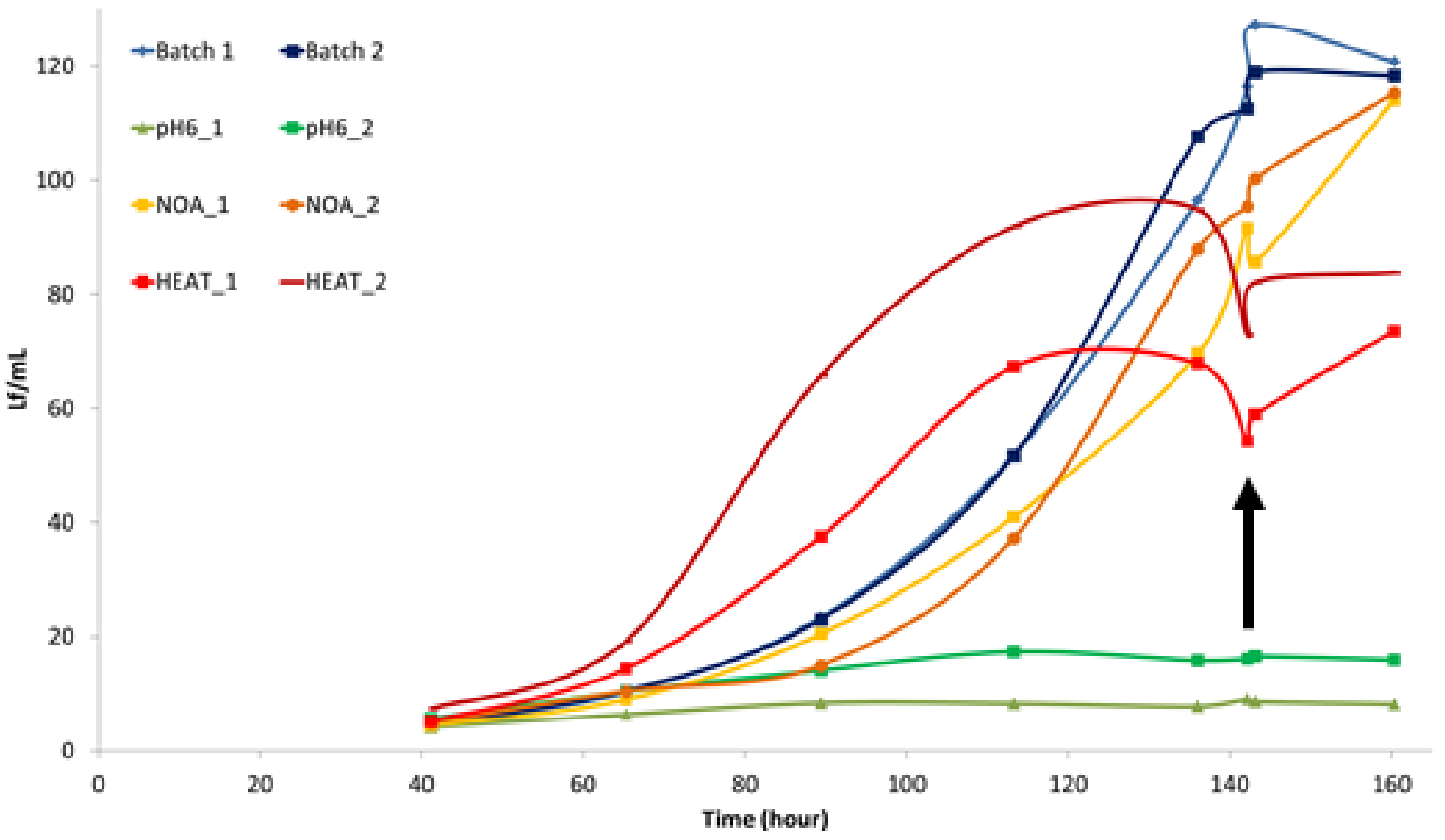

**Figure 5**

**Table 1.**

| Sequence | TeNT position (#AA) | Start and End Scan Time (min) | Transitions | Cone voltage (V) | Collision Energy (V) |
|---|---|---|---|---|---|
| GLDIYYK | Light chain 29-35 | 4.2 to 5.8 | 436.40 > 701.30 ( $y^{5+}$ as quantifier)<br>436.40 > 586.40 ($y^{4+}$ as qualifier) | 30<br>30 | 15.0<br>15.0 |
| GLD(I*)YYK | IS | 4.2 to 5.8 | 439.80 > 708.20 ( $y^{5+}$ as quantifier)<br>439.80 > 593.30 ($y^{4+}$ as qualifier) | 30<br>30 | 12.0<br>16.0 |
| VGYNAPGIPLYK | Heavy chain 1227-1238 | 5.2 to 6.8 | 646.4 > 787.40 ( $y^{7+}$ as quantifier)<br>646.4> 633.40 ($y^{5+}$ as qualifier) | 30<br>30 | 19.0<br>18.0 |
| VGYNAPG(I*)PLYK | IS | 5.2 to 6.8 | 649.90 > 794.30 ($y^{7+}$ as quantifier)<br>649.9 > 640.40 ($y^{5+}$ as qualifier) | 30<br>30 | 21.0<br>18.0 |
| VAYQAPGIPLYKGLEIYYK | ISd | 6.2 to 7.4 | 729.40 > 827.00 ($y^{14++}$ as qualifier) | 45 | 25 |
| VAYQAPGIPLYK | Isd fragment | 5.2 to 6.8 | 660.40 > 858.60 ($y^{8+}$ as qualifier) | 45 | 21 |

**Table 2.**

| | | Concentration Lf/mL | Lc peptide as quantifier | Hc peptide as quantifier | average of both peptides | Predefined acceptance criteria |
|---|---|---|---|---|---|---|
| Linearity ($R^2$) | | / | 0,999 | 0,999 | 0,999 | 0.99 |
| trueness % | | 0,25 | +2,22 | +4,44 | +3,33 | ± 15 %, except at LLOQ where ± 20 % is allowed |
| | | 1,25 | +2,56 | +2,38 | +2,47 | |
| | | 6,25 | +8,63 | +10,46 | +9,54 | |
| | | 18,75 | +2,89 | +4,44 | +3,66 | |
| | | 25 | +7,51 | +8,46 | +7,99 | |
| Precision | repeatability % | 0,25 | 12,63 | 11,35 | 10,57 | 15 %, except at LLOQ where 20 % is allowed |
| | | 1,25 | 9,18 | 11,93 | 8,69 | |
| | | 6,25 | 4,45 | 5,33 | 4,30 | |
| | | 18,75 | 5,36 | 7,93 | 5,91 | |
| | | 25 | 7,93 | 6,38 | 6,20 | |
| | intermediate precision % | 0,25 | 12,63 | 12,61 | 10,57 | 15 %, except at LLOQ where 20 % is allowed |
| | | 1,25 | 11,72 | 11,93 | 9,71 | |
| | | 6,25 | 8,22 | 7,79 | 7,14 | |
| | | 18,75 | 7,95 | 10,04 | 8,41 | |
| | | 25 | 8,54 | 6,50 | 6,24 | |
| Method accuracy | | 0,25 | [-23,77; 28,22] | [-22,11; 30,99] | [-18,43; 25,10] | [-30 %; + 30 %], except at LLOQ where [-40 %; +40 %] is allowed |
| | | 1,25 | [-23,10; 28,21] | [-22,17; 26,93] | [-17,99; 22,92] | |
| | | 6,25 | [-11,27; 28,52] | [-7,26; 28,17] | [-7,29; 26,37] | |
| | | 18,75 | [-15,28; 21,06] | [-17,47; 26,35] | [-15.32; 22,65] | |
| | | 25 | [-10,32; 25,34] | [-4,96; 21,89] | [-4,87; 20,85] | |
| Uncertainty | relative expanded uncertainty % | 0,25<br>1,25<br>6,25<br>18,75<br>25 | 25,61<br>24,39<br>17,45<br>16,70<br>17,47 | 25,90<br>24,18<br>16,34<br>20,87<br>13,22 | 21,44<br>19,94<br>15,10<br>17,62<br>12,66 | n.d. |

## Supplemental table 1:

| Sequences | Results Blastp | | Light chain (Lc) or heavy chain (Hc) | lowest digestion probability (%) | MS compatibility | | Most MS responsive and stable peptide of each chain |
|---|---|---|---|---|---|---|---|
| | Sequence found only in *Cl. tetani:* | Sequence specific for TeNT: | | | number of AA | MS compatibility | |
| **GLDIYYK** | ✓ | ✓ | Lc | 100 | 7 | ✓ | ✓ |
| **IIPPTNIR** | ✓ | ✓ | Lc | 98.5 | 8 | ✓ | |
| TTPVTK | ✓ | capsid protein | | | | | |
| **IYSYFPSVISK** | ✓ | ✓ | Hc | 100 | 11 | ✓ | |
| **IYSGPDK** | ✓ | ✓ | Hc | 83.8 | 7 | ✓ | |
| **FIGITELK** | ✓ | ✓ | Hc | 82.6 | 8 | ✓ | |
| **VFSTPIPFSYSK** | ✓ | ✓ | Hc | 100 | 12 | ✓ | |
| DSAGEVR | ✓ | ✓ | Hc | 100 | 7 | very hydrophilic peptide | |
| **SGDFIK** | ✓ | ✓ | Hc | 100 | 6 | ✓ | |
| **VGYNAPGIPLYK** | ✓ | ✓ | Hc | 100 | 12 | ✓ | ✓ |

**Supplemental Table 2.**

| | 0.25 Lf/mL | 1.25 Lf/mL | 6.25 Lf/mL | 18.75 Lf/mL | 25 Lf/mL |
|---|---|---|---|---|---|
| Mean RSD % | 7.70 | 6.38 | 3.44 | 4.05 | 4.64 |
| Median RSD% | 6.73 | 4.40 | 2.25 | 4.39 | 4.60 |
| Maximum RSD% | 18.44 | 19.41 | 9.39 | 7.04 | 10.91 |

**Supplemental Table 3.**

[TeNT] in supernatants expressed in Lf/mL

| time (h) | Batch 1 | | Batch 2 | | pH6_1 | | pH6_2 | | NOA_1 | | NOA_2 | | HEAT_1 | | HEAT_2 | |
|---|---|---|---|---|---|---|---|---|---|---|---|---|---|---|---|---|
| | mean | stdev | mean | stdev | mean | stdev | mean | stdev | mean | stdev | mean | stdev | mean | stdev | mean | stdev |
| 41 | 5.0 | 0..8 | 4.6 | 0.2 | 4.2 | 0.2 | 5.6 | 0.7 | 4.5 | 0.2 | 4.9 | 0.2 | 5.2 | 0.6 | 7.4 | 0.4 |
| 65 | 10.2 | 0.5 | 10.4 | 1.1 | 6.4 | 0.6 | 10.3 | 0.2 | 9.0 | 0.8 | 10.4 | 0.1 | 14.4 | 1.3 | 19.2 | 2.4 |
| 89 | 23.4 | 0.5 | 23.0 | 1.8 | 8.4 | 0.3 | 14.1 | 1.0 | 20.5 | 1.6 | 15.0 | 0.3 | 37.6 | 5.7 | 65.8 | 7.2 |
| 113 | 51.8 | 5.6 | 51.7 | 4.4 | 8.3 | 0.6 | 17.3 | 0.4 | 41.0 | 8.3 | 37.3 | 0.9 | 67.3 | 2.3 | 91.9 | 5.1 |
| 136 | 96.5 | 15.7 | 107.6 | 24.7 | 7.8 | 0.7 | 15.8 | 0.5 | 69.6 | 0.1 | 87.9 | 0.1 | 67.9 | 5.4 | 94.8 | 12.1 |
| 142 | 116.4 | 4.9 | 112.5 | 20.9 | 9.1 | 0.1 | 16.1 | 0.4 | 91.6 | 2.7 | 95.4 | 1.6 | 54.3 | 2.4 | 73.0 | 4.9 |
| 143 | 127.2 | 29.2 | 118.9 | 8.4 | 8.6 | 0.4 | 16.5 | 4.0 | 85.8 | 8.3 | 100.3 | 2.3 | 58.9 | 2.5 | 82.1 | 4.9 |
| 160 | 120.8 | 14.2 | 118.3 | 17.5 | 8.2 | 0.2 | 15.9 | 0.4 | 114.0 | 6.9 | 115.3 | 19.0 | 73.6 | 10.0 | 83.8 | 9.9 |

## FUNDING

This work has been funded by the EU/EFPIA/Innovative Medicines Initiative 2 Joint Undertaking, termed the VAC2VAC project (grant agreement N-115924). The funding source was not involved in the design of the study, analysis and interpretation of the data, nor the writing and submission of this manuscript.

## CONFLICTS OF INTEREST

Raphaël Esson, Eric Abachin, Bruce Carpick and Sylvie Uhlrich, are all employees of Sanofi Pasteur and may hold shares and/or stock options in the company. Jean-Francois Dierick is an employee of the GSK group of companies. Antoine Francotte, Melissa Vanhamme, Alexandre Dobly and Celine Vanhee, have no conflicts of interest.

## ACKNOWLEDGEMENTS

We would like to thank Dr. Laurent Thion, employed at the initiation of this project by Sanofi Pasteur, for the fruitful scientific discussions.